\documentclass[11pt,fleqn]{article}
\usepackage{epsfig}
\def\greaterthansquiggle{\raise.3ex\hbox{$>$\kern-.75em\lower1ex\hbox{$\sim$}}}
\def\lessthansquiggle{\raise.3ex\hbox{$<$\kern-.75em\lower1ex\hbox{$\sim$}}}
\newcommand{\la}{\label}
\newcommand{\re}{\ref}
\newcommand{\ci}{\cite}
\newcommand{\beqn}{\begin{eqnarray}}
\newcommand{\eeqn}{\end{eqnarray}}
\newcommand{\bequ}{\begin{equation}}
\newcommand{\eequ}{\end{equation}}
\newcommand{\bsl}{\begin{sloppypar}}
\newcommand{\esl}{\end{sloppypar}}
\newcommand{\grts}{\greaterthansquiggle}
\newcommand{\lets}{\lessthansquiggle}

\begin{document}
\null
\hfill LC-TH-2000-033\\
\vspace{.1cm}\hfill DESY 00-003\\
\vspace{.1cm}\hfill UWThPh-2000-4\\
\vspace{.1cm}\hfill WUE-ITP-2000-005\\
\vspace{.1cm}\hfill HEPHY-PUB 727/2000\\
\vspace{.1cm}\hfill hep-ph/0004181\\
\vskip .4cm

\begin{center}
{\Large \bf Beam Polarization and Spin Correlation Effects\\[.5em]
in Chargino Production and Decay
\footnote{Contribution to the Proceedings of the 2nd Joint ECFA/DESY study on
Physics\\\mbox{\hspace{.5cm}} and Detectors for a Linear Electron-Positron 
Collider}%
}
\end{center}
\vskip 1em
{\large
{\sc G.~Moortgat--Pick$^{a}$\footnote{e-mail:
    gudrid@mail.desy.de},
A.~Bartl$^{b}$\footnote{e-mail:
     bartl@ap.univie.ac.at},  
H. Fraas$^{c}$\footnote{e-mail:
    fraas@physik.uni-wuerzburg.de},
W.~Majerotto$^{d}$\footnote{e-mail:
     majer@hephy.oeaw.ac.at.}%
}}\\[1ex]
{\footnotesize \it
$^{a}$ DESY, Deutsches Elektronen--Synchrotron, D-22603 Hamburg, Germany}\\
{\footnotesize \it
$^{b}$ Institut f\"ur Theoretische Physik, Universit\"at Wien, A-1090 Wien, 
Austria}\\
{\footnotesize \it
$^{c}$ Institut f\"ur Theoretische Physik, Universit\"at
W\"urzburg, D-97074~W\"urzburg, Germany}\\
{\footnotesize \it
$^{d}$ Institut f\"ur Hochenergiephysik, \"Osterreichische
 Akademie der Wissenschaften,\\\phantom{$^{d}$} A-1050 Wien, Austria}
\vskip 1em
\par
\vskip .8cm

\begin{abstract}
We study chargino production  
$e^+ e^-\to \tilde{\chi}^+_1 \tilde{\chi}^-_1$ and the subsequent leptonic 
decay $\tilde{\chi}^-_1\to \tilde{\chi}^0_1 e^- \bar{\nu}_e$ 
 including the complete spin correlations between 
production and decay. We work out the advantages of polarizing 
the $e^+$ and $e^-$ beams. 
We study in detail the polarized cross sections, 
the angular distribution and the forward--backward 
asymmetry of the decay electron. 
They can be used to determine the sneutrino mass $m_{\tilde{\nu}_e}$.
\end{abstract}

\vspace{1em}
\hfill

\section{Introduction}
If weak--scale supersymmetry (SUSY) is realized in nature, 
then SUSY particles will be observable at an $e^+ e^-$ linear collider
with c.m.s. energy in the range $\sqrt{s} \le$ 1~TeV. The
experimental study of charginos and the determination of their
properties will be particularly important. The charginos
$\tilde{\chi}_i^\pm$, $i = 1,2$, are the mass eigenstates of the charged
W-ino--higgsino system. In the Minimal Supersymmetric Standard Model
(MSSM) their mass eigenvalues and eigenstates are
determined by the parameters $M_2$, $\mu$ and $\tan \beta$.

Previous papers mainly analyzed production cross sections and decay
branching ratios in the MSSM (see, e.g. \ci{bartl,Ambrosanio} and references 
therein). Recently in \ci{Choi} a  
method has been proposed for determining the SUSY parameters $M_2$, $\mu$,
$\tan\beta$ and $m_{\tilde{\nu}_e}$ by measuring suitable observables in 
$e^+ e^-\to \tilde{\chi}^+_i \tilde{\chi}^-_j$, $i,j=1,2$.
For a determination of the properties of charginos it is necessary to measure
also angular distributions
and angular correlations of the decay products.
In the calculation of these observables the full spin correlations
between chargino production and decay have to be taken into account.
This has been done in recent analyses in \ci{Gudi_char,Lafage}, 
where the process 
\bequ
e^+ e^- \to \tilde{\chi}_i^+ \tilde{\chi}_j^- \la{eq_0a}
\eequ
with polarized beams and the subsequent leptonic decays
\bequ
\tilde{\chi}_j^- \to \tilde{\chi}_k^0 \ell^- \bar{\nu}_\ell
\la{eq_0b}
\eequ
have been studied.
In \ci{Gudi_char} 
analytical formulae including the complete spin correlations have been
presented. These formulae are also applicable in the case of complex
couplings. The Feynman diagrams for the production (\re{eq_0a})
and the decay (\re{eq_0b}) are shown in Fig.~\re{feyn}.

The present paper is based on our analyses in 
\ci{Gudi_char,Gudi_sneu}. We study
chargino production and decay with polarized $e^+$ and $e^-$ beams.
We use the MSSM as general framework and give predictions for the
total cross section, the angular distribution of the decay lepton, and
its forward--backward asymmetry. We show that using polarized $e^+$ beams 
together with polarized $e^-$ beams can enhance the cross sections and
can give additional information on the mixing character of the charginos
and the masses of the exchanged particles. We study
the $m_{\tilde{\nu}_e}$ and $m_{\tilde{e}_L}$ dependence of the decay lepton
forward--backward asymmetry, which can be used to determine 
$m_{\tilde{\nu}_e}$.
\section{Differential Cross Section}
The differential cross section 
for the combined reaction of (\re{eq_0a}) and (\re{eq_0b}) is given by:
\begin{equation}
d\sigma_e=\frac{1}{2s}|T|^2 (2\pi)^4
\delta^4(p_1+p_2-\sum_{i} p_i) d{\rm lips}(p_3\ldots p_{10})\label{eq_13},
\end{equation}
where  
$d\mbox{lips}(p_3 \ldots p_{10})$ is the Lorentz invariant phase 
space element. The amplitude squared is \ci{Gudi_char}
\beqn
|T|^2& =&4|\Delta(\tilde{\chi}_i^+)|^2|
\Delta(\tilde{\chi}_j^-)|^2
         \Big(P(\tilde{\chi}_i^+ \tilde{\chi}_j^-) 
D(\tilde{\chi}_i^+) D(\tilde{\chi}_j^-)\nonumber\\
&& +\sum^3_{a=1}\Sigma_P^a(\tilde{\chi}_i^+) 
\Sigma_D^a(\tilde{\chi}_i^+) 
D(\tilde{\chi}_j^-)
+\sum^3_{b=1}\Sigma_P^b(\tilde{\chi}_j^-) \Sigma_D^b(\tilde{\chi}_j^-)
D(\tilde{\chi}_i^+)\nonumber\\&&
    +\sum^3_{a,b=1}\Sigma_P^{ab}(\tilde{\chi}_i^+\tilde{\chi}_j^-)
 \Sigma^a_D(\tilde{\chi}_i^+) 
\Sigma^b_D(\tilde{\chi}_j^-)\Big),
\la{eq4_5}
\eeqn
where $P(\tilde{\chi}_i^+ \tilde{\chi}_j^-)$ ($D(\tilde{\chi}_i^+)$, 
$D(\tilde{\chi}_j^-)$ ) denotes the part of the unnormalized
spin density matrix 
(decay matrix), which is independent of the chargino polarization, and
$\Sigma_P^a(\tilde{\chi}_i^+)$, $\Sigma_P^b(\tilde{\chi}_j^-)$,
$\Sigma_P^{ab}(\tilde{\chi}_i^+\tilde{\chi}_j^-)$ 
($\Sigma_D^a(\tilde{\chi}_i^+)$, $\Sigma_D^b(\tilde{\chi}_j^-)$) denote
those parts of the spin density matrix (decay matrix) which depend on the 
chargino polarization.
If all spin correlations are neglected, then only the first term
contributes. The second and third term describe the spin correlations
between production and decay. $\Sigma_p^1$ denotes the transverse
polarization in the production plane, $\Sigma_p^2$ is the polarization
perpendicular to the production plane, and $\Sigma_p^3$ is the longitudinal
polarization of the decaying chargino. $\Sigma_P^{ab}$ is due to the
correlations between the polarizations of both decaying charginos,
with $a,b=1,2$ referring to transverse chargino
polarizations, and $a,b=3$ referring 
to longitudinal polarization. The chargino propagators are given by
$\Delta(\tilde{\chi}^{\pm}_{k})=1/[p_k^2-m_{k}^2+i m_{k} 
\Gamma_{k}]$ with four--momentum $p_{k}$, mass $m_{k}$, and
total width $\Gamma_{k}$ of the decaying particle.
For more details we refer to \ci{Gudi_char,Gudi_sneu,LC99_neut}.
\section{Numerical Analysis and Discussion}\la{sec:2}
In the MSSM \ci{Haber-Kane} the masses and couplings of
 charginos and neutralinos are 
functions of the parameters $M_1$, $M_2$, $\mu$, $\tan\beta$, with $M_1$ 
normally fixed by the GUT relation $M_1=\frac{5}{3} M_2 \tan^2\Theta_W$. As we
do not consider CP violating effects, the parameters can be chosen real. The 
explicit expressions for the neutralino and chargino mass mixing matrices
can be found in \ci{bartl} (note that in 
Refs.~\ci{bartl,Gudi_char,Gudi_sneu} the notation 
$M'$ and $M$ for $M_1$ and $M_2$ was used).

We will study a gaugino--like and a higgsino--like scenario, which
we denote bei A and B, respectively.  
Scenario~A \ci{Blair} is a mSUGRA scenario with the corresponding MSSM 
parameters given in Table~\ref{tab:1}.
$\tilde{\chi}^{\pm}_1$ and $\tilde{\chi}^0_1$ have a large gaugino 
component. 
For easier comparison we want to have in the higgsino--like scenario~B
similar masses for $\tilde{\chi}^{\pm}_1$, $\tilde{\chi}^0_1$,
$\tilde{\nu}_e$ and $\tilde{e}_L$.
Therefore we have relaxed the GUT relation for the gaugino mass parameters.
The parameters of scenario~B are given in Table~\re{tab:1}. 
\subsection{Beam polarization dependence of the total cross section}\la{sec:21}
\bsl
In Figs.~\re{fig_1}a and b we show the cross section 
$\sigma(e^+ e^-\to\tilde{\chi}^+_1 \tilde{\chi}^-_1)$ at
$\sqrt{s}=2 m_{\tilde{\chi}^{\pm}_1}+10$~GeV for 
scenarios~A and B, 
respectively, as a function of the electron beam 
polarization $P_-^3$ and positron beam polarization $P_+^3$ (with
$P^3_{\pm}=\{-1, 0, 1\}$ for $\{$left--,un--, right--$\}$ polarized).
The white area is covered by an electron 
polarization $|P_-^3|\le 85\%$ and a positron polarization $|P_+^3|\le 60\%$.
It can be seen that the beam polarizations may be used to enhance the cross 
section. One can gain a factor of about 
two by polarizing the positron beam in 
addition to the electron beam. Owing to the $\tilde{\nu}_e$ exchange 
the effect is biggest if the electron is left and the positron right polarized.
We choose $\sqrt{s}$ not too far from threshold, because the spin 
correlations to be discussed below
decrease with $\sqrt{s}$ \ci{Gudi_char}.
\esl

Since in the chargino process the $\tilde{\nu}_e$ exchange in the t--channel 
favours left polarized electron beams and right polarized
positron beams, one expects for gaugino--like scenarios 
the following sequence of polarized cross sections \ci{Gudi_diss} for 
$|P_-^3|=85$\% and $|P_+^3|=60$\% ( these values of $|P_-^3|$ and $|P_+^3|$ 
are used throughout the paper): 
\bequ
\sigma_e^{-+}>\sigma_e^{-0}>\sigma_e^{00}>\sigma_e^{--}>\sigma_e^{++}>
\sigma_e^{+0}>\sigma_e^{+-}.
\la{eq_gauge}
\eequ
Here $(-+)$ etc. denotes the sign of the electron polarization $P_-^3$ and 
of the positron polarization $P_+^3$, respectively, and $\sigma_e$ is 
defined as
\bequ
\sigma_e=\sigma(e^+ e^-\to \tilde{\chi}^+_1 \tilde{\chi}^-_1)
\times BR(\tilde{\chi}^-_1\to \tilde{\chi}^0_1 e^- \bar{\nu}_e),
\la{sigma_e}
\eequ
where we assume that only one chargino decays leptonically.

For pure higgsinos near threshold one would have due to $Z^0$ exchange
\bequ
\sigma_e^{-+}>\sigma_e^{+-}>\sigma_e^{-0}>
\sigma_e^{00}>\sigma_e^{+0}>\sigma_e^{--}>\sigma_e^{++}.
\la{eq_higgs}
\eequ
Since in chargino pair production also $\gamma$ exchange contributes, the 
relations
(\re{eq_gauge}), (\re{eq_higgs}) are only approximately valid.
Nevertheless, one can get additional information if  
the electron and the positron beam are polarized, since the sequences of 
polarized cross sections for gaugino--like and higgsino--like 
scenarios are different. 
If one had only the electron beam polarized, in both scenarios one would 
obtain the same sequence of polarized cross sections, namely 
$\sigma^{-0}>\sigma^{00}>\sigma^{+0}$.

In scenario~A we get for $|P_-^3|=85\%$ and 
$|P_+^3|=60\%$ at 
$\sqrt{s}=2m_{\tilde{\chi}^{\pm}_1}+10$~GeV the results
given in Table~\re{table_11}.
They fulfil the relation (\re{eq_gauge}) for polarized cross sections.
However, in the higgsino--like scenario~B the sequence
is different from relation (\re{eq_higgs}) 
due to $\gamma$ exchange and in particular  
$\gamma Z^0$ interference.
\subsection{Spin effects in the lepton angular distribution and in the lepton 
forward--backward asymmetry}
We will discuss the lepton angular distribution $d\sigma_e/d\cos\Theta_e$
in the overall c.m.s. of the combined reactions (\re{eq_0a}) and (\re{eq_0b}). 
Here $\Theta_e$ denotes
the angle between the electron beam and the outgoing $e^-$.
The forward--backward asymmetry $A_{FB}$ of the decay lepton is
defined as
\bequ
A_{FB}=\frac{\sigma_e(\cos\Theta_e>0)-\sigma_e(\cos\Theta_e<0)}
{\sigma_e(\cos\Theta_e>0)+\sigma_e(\cos\Theta_e<0)}. \la{eq_22}
\eequ
This observable is very sensitive to the gaugino component of the chargino and 
the mass of the exchanged sneutrino and slepton.
While in $\sigma_e$, eq.~(\re{sigma_e}), the leptonic branching ratio 
of $\tilde{\chi}^-_1$ enters, which depends on the parameters of the squark 
sector, $A_{FB}$, eq. (\re{eq_22}), has the advantage of being independent 
of the squark sector.

First we discuss how the spin correlations depend on the mixing character 
of the charginos \ci{Gudi_char}. 
In Figs.~\re{fig_2}a and b we show the  angular 
distribution $d\sigma_e/d\cos\Theta_e$ of the decay lepton
for scenarios~A  and ~B, respectively, for unpolarized beams and for both 
beams polarized at 
$\sqrt{s}=2 m_{\tilde{\chi}^{\pm}_1}+10$~GeV. 
The angular distributions 
are compared with those, where no spin correlations are taken into account.
Without spin correlations the $\cos\Theta_e$ dependence would be much flatter. 
It can also be seen 
that the spin correlations are more important in the gaugino--like scenario.

In Figs.~\re{fig_3}a and b 
$A_{FB}$ is shown with and without spin correlations
as a function of $\sqrt{s}$ for 
the two scenarios. 
In the gaugino--like scenario~A  $A_{FB}$ can reach
40\% at $\sqrt{s}=500$~GeV.
The large asymmetry in scenario~A, 
Fig.\re{fig_3}a, is due to the $\tilde{\nu}_e$ exchange in the crossed channel.
Also the $\tilde{e}_L$ exchange in the decay
$\tilde{\chi}^-_1\to \tilde{\chi}^0_1 e^- \bar{\nu}_e$ plays an important
role. Therefore in the gaugino--like scenario~A $A_{FB}$ depends  
appreciably on $m_{\tilde{e}_L}$ (see also Fig.~\re{fig_5}).

In the higgsino--like scenario, Fig.~\re{fig_3}b, $Z^0$ and
$W^{\pm}$ exchanges in production and decay, 
respectively,
dominate. Therefore, $A_{FB}$ is much smaller, 
$A_{FB}\le 8\%$. 
However, in both scenarios, close to threshold,
$A_{FB}$ would be one order of magnitude smaller if the spin 
correlations are neglected.
For fixed chargino masses 
the correlations decrease with $\sqrt{s}$. 
This happens more rapidly in the gaugino case. 
At energies far from the threshold the charginos 
have a large energy, and the decay lepton has essentially the same 
direction as the chargino \ci{Feng}.

In scenario~A the dependence of $A_{FB}$ on the polarizations
of the beams is very weak, see Table~\re{table_11}.
The enhancement of the cross section 
for left polarized electron beams and for right polarized positron beams
is cancelled in
the ratio of the cross sections $\sigma_e(\cos\Theta_e)$
in (\re{eq_22}). 
In scenario~B $A_{FB}$ depends more strongly
on the polarizations of the beams, see Table \re{table_11}.
The $Z^0$ couplings to the higgsino components of the 
charginos determine the behaviour of $A_{FB}$. The sign of $A_{FB}$ can even
flip if 
the polarization of the electron beam is changed from left to right due to 
the dominating axial coupling of $Z^0 e^+ e^-$. 
However, since the crossed channel contributions 
are suppressed, $A_{FB}$ is smaller than in scenario~A and lies 
between $-6\%$ and +8\%. 
\begin{table}[h]
\begin{center}
\begin{tabular}{|l||c|c|c|c|c|c|c|c|c|c|c|}
 & $M_1$ & 
$M_2$ & $\mu$ & $\tan\beta$ & $m_{\tilde{e}_L}$ & $m_{\tilde{\nu}_e}$ & 
$m_{\tilde{\chi}^{\pm}_1}$ &
$m_{\tilde{\chi}^{\pm}_2}$ & $m_{\tilde{\chi}^{0}_1}$ & 
$m_{\tilde{\chi}^0_2}$ &
$\Gamma_{\tilde{\chi}^{\pm}_1}$ \\ \hline
 A & 78 & 152 & 316 & 3 & 176 & 161 & $128$ & 357 & 71 & 130 & 84E$-6$ \\\hline
B & 95 & 400 & 145 & 3 & 176 & 161 & 129 & 421 & 71 & $149$ &  217E$-6$
\\ \hline
\end{tabular}
\caption{SUSY parameters and masses in scenarios~A \cite{Blair} and B. 
Masses and total width are given in GeV. \label{tab:1}}
%
\begin{tabular}{|l||c|c|c|c|c|c|c|}
 & \multicolumn{7}{|c|}
{$\sqrt{s}=2 m_{\tilde{\chi}^{\pm}_1}+10$~GeV} \\ \hline
A & $(-+)$ & $(-0)$ &$(00)$ &$(--)$ & $(++)$ &$(+0)$ & $(+-)$\\ 
\hline
 $\sigma_e$/fb & 59 & 37 & 20 &15  & 5  & 3 & 1 \\ \hline
$A_{FB}$/\% & 33 & 33 & 33 & 33 & 32 & 31 & 25 \\ \hline\hline
B & $(-+)$ & $(-0)$ &$(00)$ &$(--)$ & $(+-)$ &$(+0)$ & $(++)$\\ 
\hline
 $\sigma_e$/fb & 152  & 96 & 59 & 40 & 26   & 22   & 18 \\ \hline
$A_{FB}$/\%    & 8    & 8  & 6  & 7  & $-6$ & $-3$ & 3 \\ \hline
\end{tabular}
\caption{Polarized cross sections 
$\sigma_e=\sigma(e^+ e^-\to \tilde{\chi}^+_1 \tilde{\chi}^-_1)
\times BR(\tilde{\chi}^-_1\to \tilde{\chi}^0_1 e^- \nu_e)$/fb
and forward--backward asymmetries $A_{FB}$ of the decay electron 
at $\sqrt{s}=2 m_{\tilde{\chi}^{\pm}_1}+10$~GeV
in scenarios~A and B, see Table~\re{tab:1}, for unpolarized beams 
$(00)$, only 
electron beam polarized $(-0)$, $(+0)$ with  $P_-^3=\pm 85\%$
and both beams polarized with 
$P_-^3=\pm 85\%$, $P_+^3=\pm 60\%$. \la{table_11}}
\end{center}
\end{table}
\subsection{Sneutrino mass dependence of $\sigma_e$ and $A_{FB}$}\la{sec:23}
If the chargino $\tilde{\chi}^{\pm}_1$ has a substantial gaugino 
component, the sneutrino exchange in the t-channel has a strong
influence on the cross section and angular distribution of
chargino production. In \cite{fpmt,Tsukamoto} it was
studied if the sneutrino mass $m_{\tilde{\nu}_e}$ can be
determined from the angular distribution of the production process $e^+e^- \to
\tilde{\chi}^+_1 \tilde{\chi}^-_1$. In this subsection we study the
$m_{\tilde{\nu}_e}$ dependence of the cross section $\sigma_e$, 
eq.~(\re{sigma_e}), as well as the decay lepton angular distribution of 
$e^+e^- \to \tilde{\chi}^+_1 \tilde{\chi}^-_1$, $\tilde{\chi}^-_1 \to
\tilde{\chi}^0_1 e^- \bar{\nu}_{e}$, and the decay lepton
forward--backward asymmetry $A_{FB}$, eq.~(\re{eq_22}). As these
observables depend decisively on the spin correlations, it is
instructive to have a closer look on their $m_{\tilde{\nu}_e}$
dependence. These observables also depend on the slepton mass 
$m_{\tilde{e}_L}$, due to the $\tilde{e}_L$ exchange in the
decay amplitude. Since $\tilde{\ell}_L$ and $\tilde{\nu}_{\ell}$ are
members of the same $SU(2)_L$ doublet, we assume the relation
\cite{Tsukamoto,Ramond} 
\bequ
m^2_{\tilde{\ell}_L} = m^2_{\tilde{\nu}_{\ell}} - m^2_W \cos2\beta \la{eq_su2}
\eequ
between their masses, with $m_W$ the mass of the $W^{\pm}$
boson. As this relation is based on weak $SU(2)_L$ symmetry, it
has to be fulfilled at tree level, and can only be modified by
radiative corrections.

In Figs.~\re{fig_4}a and b we show $\sigma_e$ as a function of 
$m_{\tilde{\nu}_e}$ for three sets of $e^-$ and $e^+$ beam
polarizations $(P^3_-,P^3_+)=(-85\%,+60\%)$, $(-85\%,0)$,
$(0,0)$ at $\sqrt{s} =2 m_{\tilde{\chi}^{\pm}_1} + 10$~GeV and 
$\sqrt{s} = 500$~GeV, respectively. The other SUSY parameters, 
apart from $m_{\tilde{\nu}_e}$ and $m_{\tilde{e}_L}$, are as 
in scenario A (Table~\re{tab:1}). 
The cross section $\sigma_e$, as shown in Fig.~\re{fig_4}a and b, exhibits
a pronounced minimum, which is due to the destructive interference between 
$Z$ exchange and $\tilde{\nu}_e$ exchange.
For $\sqrt{s}$ near threshold this minimum is
approximately at $m_{\tilde{\nu}_e} \approx m_{\tilde{\chi}^{\pm}_1}$
and in the limit $\sqrt{s} \to 2m_{\tilde{\chi}^{\pm}_1}$ 
the minimum reaches exactly $m_{\tilde{\nu}_e} \to m_{\tilde{\chi}^{\pm}_1}$. 
For $\sqrt{s}=500$~GeV the minimum is shifted to 
$m_{\tilde{\nu}_e} \approx 250$~GeV (Fig.~\re{fig_4}b).
Due to this minimum one gets an ambiguity when determining $m_{\tilde{\nu}_e}$
by measuring $\sigma_e$. 

The cross section is biggest for left--polarized $e^-$ beam and
right--polarized $e^+$ beam.
Compared to the unpolarized
cross section one can gain a factor of $1.8$ if only the $e^-$
beam is polarized $(P^3_-=-85\%)$, and a factor of about $3$ if
both beams are polarized $(P^3_-=-85\%, P^3_+=+60\%)$. Also the 
$m_{\tilde{\nu}_e}$ dependence is strongest for
$(P^3_-,P^3_+)=(-,+)$. 
For $m_{\tilde{\nu}_e}\le m_{\tilde{\chi}^{\pm}_1}$ the 
two--body decay $\tilde{\chi}^-_1 \to e^- \bar{\tilde{\nu}}_e$ is opening.

Also the decay lepton angular distribution depends significantly
on $m_{\tilde{\nu}_e}$. In Figs.~\re{fig_6}a and b we plot
the $\cos \Theta_e$ distribution at
$\sqrt{s} = 2m_{\tilde{\chi}^{\pm}_1}+10$~GeV, for unpolarized
and polarized beams, 
taking $m_{\tilde{\nu}_e}=130$~GeV and $250$~GeV, respectively. 
For $m_{\tilde{\nu}_e}=130$~GeV (Fig.~\re{fig_6}a) the $\cos\Theta_e$ 
distribution is relatively flat due to interference effects in the spin terms.
In Fig.~\re{fig_6}a, the $\cos\Theta_e$
distribution is a superposition of $Z$ exchange and $\tilde{\nu}_e$
exchange behaviour. For
$m_{\tilde{\nu}_e}=250$~GeV (Fig.~\re{fig_6}b) the $\cos\Theta_e$ distribution
is mainly due to $\tilde{\nu}_e$ exchange and therefore has its
maximum in the forward direction. 

In Figs.~\re{fig_5}a and b we show the decay lepton forward--backward
asymmetry $A_{FB}$ as a function of $m_{\tilde{\nu}_e}$ for
polarized beams at 
$\sqrt{s} = 2m_{\tilde{\chi}^{\pm}_1}+10$~GeV and 
$\sqrt{s} = 500$~GeV, respectively. In order to study the influence of 
$m_{\tilde{e}_L}$ on the decay, we calculate
$A_{FB}$ for $m_{\tilde{e}_L} = 130$~GeV, $150$~GeV, $200$~GeV, and
for $m_{\tilde{e}_L}$ fulfilling relation (\re{eq_su2}). The other SUSY
parameters are as in scenario A (Table~\re{tab:1}). The minimum of
$A_{FB}$ is due to spin correlations. 
For $\sqrt{s}=2 m_{\tilde{\chi}^{\pm}_1}+10$~GeV 
($\sqrt{s} = 500$~GeV) $A_{FB}$ depends quite
sensitively on $m_{\tilde{\nu}_e}$ up to $m_{\tilde{\nu}_e} \approx
300$~GeV ($m_{\tilde{\nu}_e} \approx 1$~TeV). The decrease of
$A_{FB}$ for $m_{\tilde{\nu}_e} \grts 200$~GeV ($m_{\tilde{\nu}_e} \grts
300$~GeV) in Fig.~\re{fig_5}a (Fig.~\re{fig_5}b) is due to the decreasing
$\tilde{\nu}_e$ exchange contribution for increasing $m_{\tilde{\nu}_e}$.
For $\sqrt{s}=2 m_{\tilde{\chi}^{\pm}_1}+10$~GeV and 
$m_{\tilde{\nu}_e}>200$~GeV $A_{FB}$ exhibits also an appreciable 
$m_{\tilde{e}_L}$ dependence 
for $m_{\tilde{e}_L}\le 200$~GeV. Since the $\tilde{e}_L$ exchange contributes
only to the decay the dependence of $A_{FB}$ on $m_{\tilde{e}_L}$ is weaker
for $\sqrt{s}=500$~GeV (Fig.~\re{fig_5}b) than near threshold.

Turning now to the question whether the sneutrino mass
$m_{\tilde{\nu}_e}$ can be determined from chargino pair production
and decay, we first consider the case that $m_{\tilde{\nu}_e} \grts
\sqrt{s}/2$, where $\tilde{\nu}_e \bar{\tilde{\nu}}_e$ pair
production is 
kinematically not possible. At $\sqrt{s}=500$~GeV the lepton
forward--backward asymmetry $A_{FB}$, as shown in Fig.~\re{fig_5}b, is
sensitive to $m_{\tilde{\nu}_e}$. 
We will estimate the precision
which can be expected if $m_{\tilde{\nu}_e}$ is determined from this
observable. We assume that the slepton mass $m_{\tilde{e}_L}$ and the
other SUSY parameters are known with good precision. For
definiteness we take, e. g., $m_{\tilde{e}_L}=200$~GeV, and the
other SUSY parameters as in scenario A. If we take 
only the statistical error $\delta(A_{FB})$ and a luminosity of
${\cal L}=500$~fb$^{-1}$, 
$A_{FB}$ can be measured up to $<\pm 1\%$. From Fig.~\re{fig_5}b we
can estimate 
that in the range $350$~GeV $\lets m_{\tilde{\nu}_e}
\lets 800$~GeV an error of about $\delta m_{\tilde{\nu}_e}<\pm 10$~GeV 
may be achieved. The experimental errors 
of $m_{\tilde{e}_L}$ and the other SUSY parameters are neglected.
The ambiguity for $\sqrt{s}=500$~GeV (Fig.~\re{fig_5}b) in the range
$250$~GeV $\lets m_{\tilde{\nu}_e} \lets 350$~GeV 
can most probably be resolved by measuring $A_{FB}$ at
different c.m.s. energies. Similarly, at
$\sqrt{s} = 2m_{\tilde{\chi}^{\pm}_1}+10$~GeV (Fig.~\re{fig_5}a), 
$A_{FB}$ is quite 
sensitive to $m_{\tilde{\nu}_e}$ in the range $130$~GeV$\lets
m_{\tilde{\nu}_e} \lets 350$~GeV where direct production is not possible. 
The ambiguity present in this $m_{\tilde{\nu}_e}$ range does not occur in 
$\sigma_e$. Hence, it can be resolved by using the
data on $A_{FB}$ and on $\sigma_e$. 
For a more quantitative assessment of the accuracy of 
$m_{\tilde{\nu}_e}$ that can be expected from measuring the decay
lepton forward--backward asymmetry in chargino production, Monte
Carlo studies taking into account experimental cuts and detector
simulation would be necessary. 

In case $m_{\tilde{\nu}_{e}} < \sqrt{s}/2$,
$\tilde{\nu}_{e} \bar{\tilde{\nu}}_{e}$ pairs can be directly
produced. If 
$m_{\tilde{\chi}^{\pm}_1} < m_{\tilde{\nu}_{e}} < \sqrt{s}/2$, then the 
visible decay $\tilde{\nu}_{e} \to e^- \tilde{\chi}^+_1$ is
kinematically allowed, and will presumably have a sufficiently
high branching ratio. We do not treat this case here, because
measuring the cross section of  
$e^+e^- \to \tilde{\nu}_{e} \bar{\tilde{\nu}}_{e}$ at threshold will
then allow us to determine $m_{\tilde{\nu}_{e}}$ with good
accuracy \cite{martyn}. If 
$m_{\tilde{\nu}_{e}} < m_{\tilde{\chi}^{\pm}_1} < \sqrt{s}/2$, then 
$\tilde{\nu}_{e}$ has no visible decay with sufficiently high
branching ratio. However, the two--body chargino decay 
$\tilde{\chi}^-_1 \to e^- \bar{\tilde{\nu}}_{e}$ is possible.
Measuring the endpoints of the energy spectrum of the decay
leptons $e^+$ and $e^-$ will provide a very precise
determination of 
the masses $m_{\tilde{\chi}^{\pm}_1}$ and $m_{\tilde{\nu}_{e}}$. 
The alternative method 
to determine $m_{\tilde{\nu}_{e}}$ by measuring the decay lepton 
forward--backward asymmetry $A_{FB}$ of chargino production
will, in principle, also be possible. However, the accuracy of
$m_{\tilde{\nu}_{e}}$ obtainable in this way is expected to be
lower than that from the decay lepton energy spectrum.
\section{Conclusions}
We have studied chargino production with both the $e^-$ and $e^+$ beam 
polarized. By an appropriate choice of the polarization of both beams one 
can obtain up to three times larger cross sections. Also the sensitivity to 
the mixing character of the charginos and to the sneutrino mass is 
considerably enhanced. By taking into account the spin correlations between
production and decay, we have investigated the angular distribution and the 
forward--backward asymmetry of the decay lepton of one of the charginos in
$e^+ e^-\to \tilde{\chi}^+_1 \tilde{\chi}^-_1$, 
$\tilde{\chi}^-_1\to \tilde{\chi}^0_1 e^- \bar{\nu}_e$. 
The forward--backward asymmetry strongly depends on the mixing character of 
the charginos. 

We have studied in detail the dependence on $m_{\tilde{\nu}_e}$. 
For appropriate 
beam polarizations the cross section is particularly 
sensitive to $m_{\tilde{\nu}_e}$. 
Measuring the angular distribution and the forward--backward asymmetry will be
useful for determining $m_{\tilde{\nu}_e}$ and thereby the mass relation
$m^2_{\tilde{\ell}_L} = m^2_{\tilde{\nu}_{\ell}} - m^2_W \cos2\beta$
can be tested. 

\vspace{-1mm}
\section*{Acknowledgements}
\vspace{-1mm}
We would like to thank U. Martyn for very valuable discussions.
We are grateful to W.~Porod and S.~Hesselbach 
for providing the computer programs for neutralino widths.
This work was also supported by 
the German Federal Ministry for
Research and Technology (BMBF) under contract number
05 7WZ91P (0), by the Deutsche Forschungsgemeinschaft under
contract Fr 1064/2-2, and the `Fonds zur
F\"orderung der wissenschaftlichen Forschung' of Austria, Project
No. P13139-PHY.

\begin{figure}[h]
\begin{minipage}{3.5cm}
\begin{center}
{\setlength{\unitlength}{1cm}
\hspace*{.7cm}
\begin{picture}(3.5,2.5)
\put(0,-1.1){\includegraphics{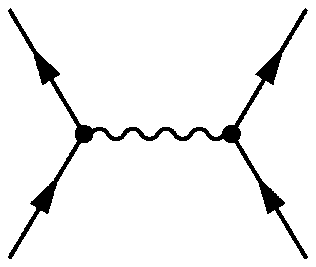}}
\put(0,-1.8){$e^{-}$}
\put(3.1,-1.8){$\tilde{\chi}^{-}_j$}
\put(0,1.7){$e^{+}$}
\put(3.1,1.7){$\tilde{\chi}^{+}_i$}
\put(1.7,.4){$\gamma$}
\end{picture}}
\end{center}
\end{minipage}
\hspace{1.5cm}
\vspace{1.6cm}
\begin{minipage}{3cm}
\begin{center}
{\setlength{\unitlength}{1cm}
\begin{picture}(3,2.5)
\put(0,-1.1){\includegraphics{prog.ps}}
\put(.2,-1.8){$e^{-}$}
\put(3,-1.8){$\tilde{\chi}^{-}_j$}
\put(.2,1.7){$e^{+}$}
\put(3,1.7){$\tilde{\chi}^{+}_i$}
\put(1.6,.4){$Z^0$}
\end{picture}}
\end{center}
\end{minipage}
\hspace{1.3cm}
\vspace{1.6cm}
\begin{minipage}{3cm}
\begin{center}
{\setlength{\unitlength}{1cm}
\begin{picture}(3,2)
\put(0,-1.5){\includegraphics{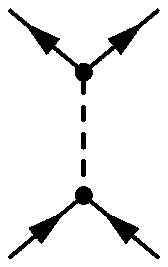}}
\put(-.3,-2){$e^{-}$}
\put(-.3,1.5){$e^{+}$}
\put(1.7,-2){$\tilde{\chi}^{-}_j$}
\put(1.7,1.5){$\tilde{\chi}^{+}_i$}
\put(.3,-.3){$\tilde{\nu}_e$}
\end{picture}}
\end{center}
\end{minipage}
\vspace{.8cm}
\begin{minipage}[h]{3cm}
\begin{center}
{\setlength{\unitlength}{1cm}
\begin{picture}(3,2.5)
\put(.8,-.1){\includegraphics{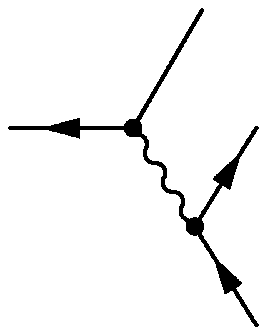}}
\put(4,-.4){$\stackrel{(-)}{\nu_{\ell}}$}
\put(.4,1.7){$\tilde{\chi}^{\pm}_{i,j}$}
\put(4,1.6){$\ell^{\pm}$}
\put(3.4,2.8){$\tilde{\chi}^0_{k,l}$}
\put(1.8,.8){$W^{\pm}$}
\end{picture}}
\end{center}
\end{minipage}
\hspace{2cm}
\vspace{.8cm}
\begin{minipage}[h]{3cm}
\begin{center}
{\setlength{\unitlength}{1cm}
\begin{picture}(3,2.9)
\put(+.3,+0){\includegraphics{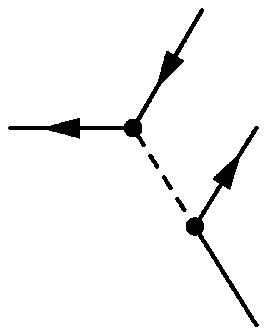}}
\put(2.6,3){$\stackrel{(-)}{\nu_{\ell}}$}
\put(3.1,2){$\ell^{\pm}$}
\put(-.3,1.8){$\tilde{\chi}^{\pm}_{i,j}$}
\put(3.1,0){$\tilde{\chi}^0_{k,l}$}
\put(1.3,1){$\tilde{\ell}_{L}$}
\end{picture}}
\end{center}
\end{minipage}
\hspace{-.2cm}
\begin{minipage}[h]{3cm}
\begin{center}
{\setlength{\unitlength}{1cm}
\begin{picture}(3,2.9)
\put(+1.5,.0){\includegraphics{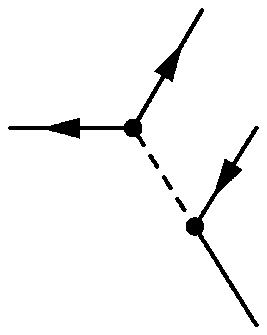}}
\put(4.2,2){$\stackrel{(-)}{\nu_{\ell}}$}
\put(3.7,3){$\ell^{\pm}$}
\put(4.4,0){$\tilde{\chi}^0_{k,l}$}
\put(.8,1.8){$\tilde{\chi}^{\pm}_{i,j}$}
\put(2.5,1){$\tilde{\nu}_{\ell}$}
\end{picture}}
\end{center}
\end{minipage}
\vspace{-1.4cm}
\caption{Feynman diagrams for the production 
$e^+e^-\to \tilde{\chi}^{+}_i \tilde{\chi}^-_j$ and 
the decay processes
$\tilde{\chi}^{\pm}_{i,j}\to \tilde{\chi}^0_{k,l} 
\ell^{\pm} \stackrel{(-)}{\nu_{\ell}}$ \la{feyn}.}
\end{figure}

\begin{figure}[t]
\hspace{-.8cm}
\begin{minipage}{7cm}
\begin{picture}(7,7)
\put(0,0){\includegraphics{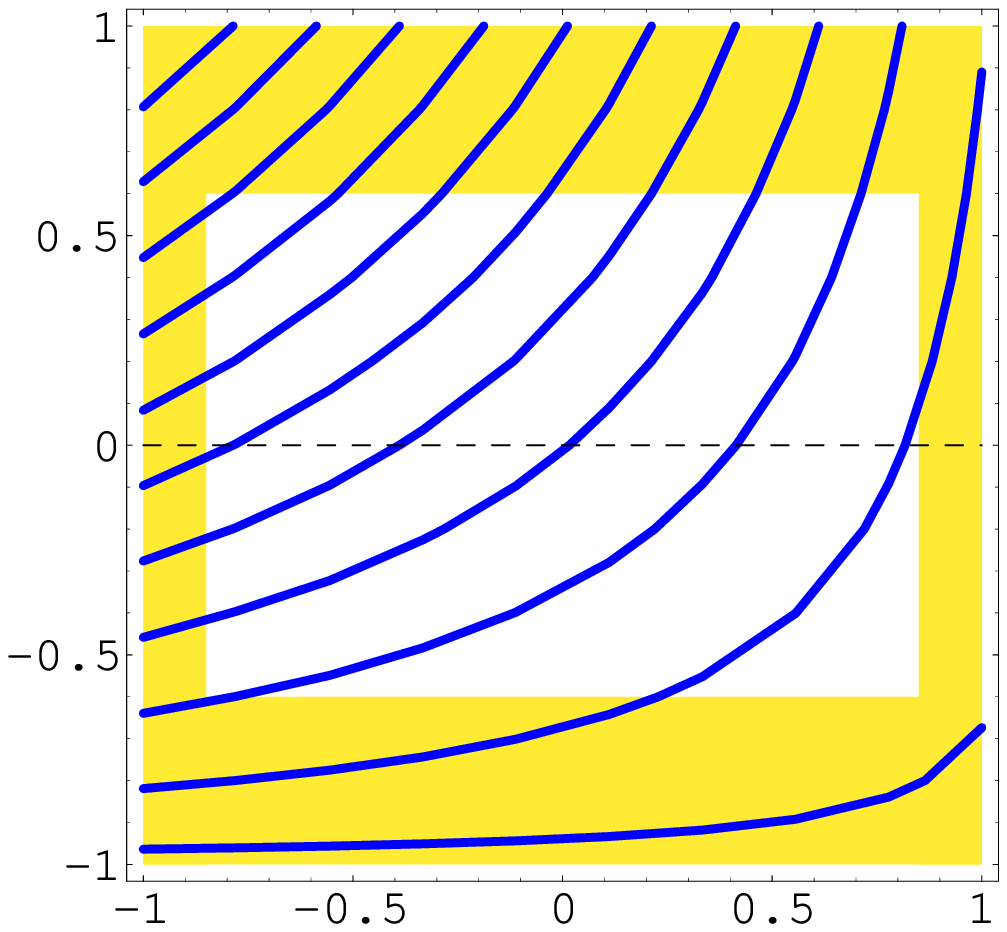}}
\put(2,4.9){a) Scenario~A}
\put(6.2,-1){\small $P_-^3$}
\put(-.1,4.6){\small $P_+^3$}
\put(5.6,-.4){\tiny 10}
\put(4.8,.6){\tiny 50}
\put(4,1.2){\tiny 100}
\put(3.4,1.7){\tiny 150}
\put(2.9,2.1){\tiny 200}
\put(2.45,2.4){\tiny 250}
\put(2.1,2.75){\tiny 300}
\put(1.8,3.05){\tiny 350}
\put(1.45,3.35){\tiny 400}
\put(1.25,3.7){\tiny 450}
\put(1.05,4.0){\tiny 500}
\end{picture}\par\vspace{.8cm}
\end{minipage}\hfill\hspace{.2cm}
\begin{minipage}{7cm}
\begin{picture}(7,7)
\put(0,0){\includegraphics{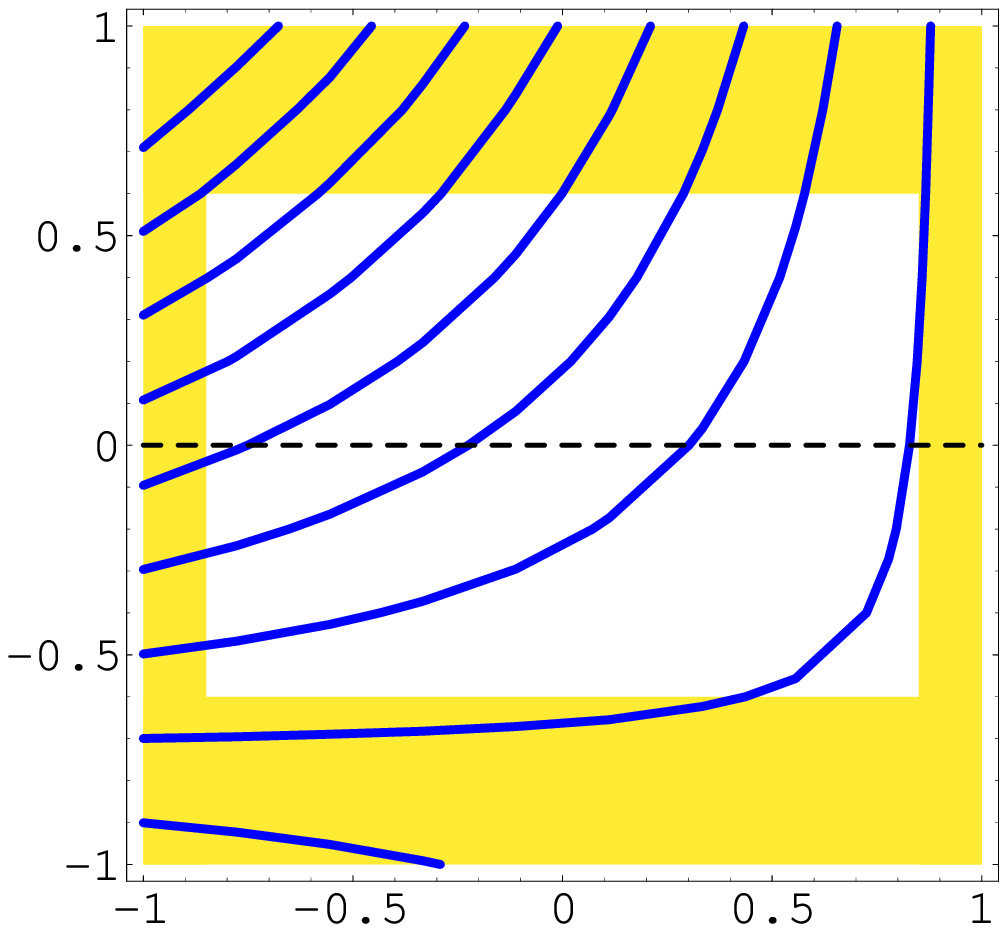}}
\put(2,4.9){b) Scenario~B}
\put(6.2,-1){\small $P_-^3$}
\put(-.1,4.6){\small $P_+^3$}
\put(1.3,-.3){\tiny 100}
\put(5.1,.5){\tiny 300}
\put(3.9,1.4){\tiny 500}
\put(3.2,2){\tiny 700}
\put(2.65,2.35){\tiny 900}
\put(2.1,2.75){\tiny 1100}
\put(1.75,3.15){\tiny 1300}
\put(1.4,3.5){\tiny 1500}
\put(1.05,3.8){\tiny 1700}
\end{picture}\par\vspace{.8cm}
\end{minipage}
\caption{Contour lines of cross 
sections $\sigma(e^+ e^-\to\tilde{\chi}^+_1\tilde{\chi}^-_1)$ 
at $\sqrt{s}=2 m_{\tilde{\chi}^{\pm}_1}+10$~GeV
in a) scenario~A and b) scenario~B.
The longitudinal 
beam polarization for electrons (positrons) is denoted by $P_-^3$ ($P_+^3$).
The white region is for $|P_{-}^3|\le 85\%$, $|P_+^3| \le 60\%$
(dashed line if only the electron beam is polarized). \la{fig_1}}
\hspace{-1cm}
\begin{minipage}{7cm}
\vspace{1.2cm}
\begin{picture}(7,5)
\put(0,0){\includegraphics{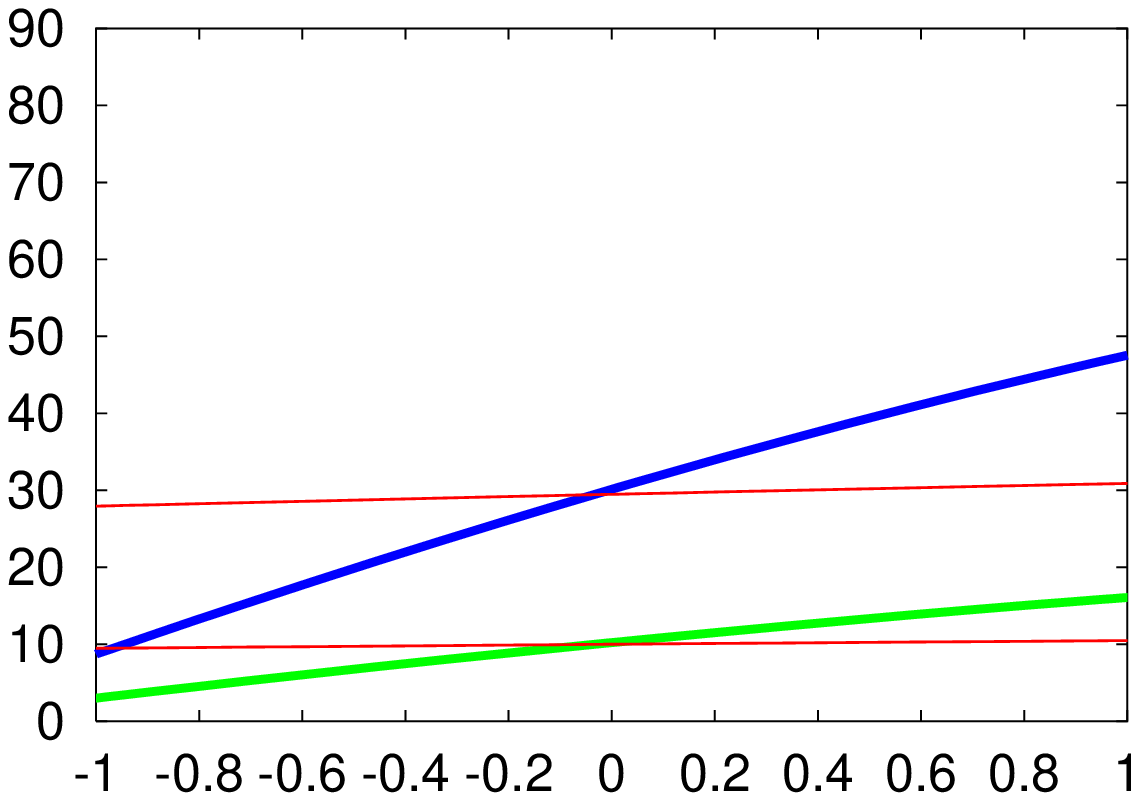}}
\put(2.3,5.1){a) Scenario~A}
\put(6.1,-.2){\small $ \cos\Theta_{e}$}
\put(-.2,5.4){\small $ \frac{d\sigma_e}{d\cos\Theta_{e}}$/fb}
\put(6,1.5){\small $(00)$}
\put(5.9,3.05){\small $(-+)$}
\end{picture}\par\vspace{.1cm}
\end{minipage}\hfill\hspace{.2cm}
\begin{minipage}{7cm}
\vspace{1.2cm}
\begin{picture}(7,5)
\put(0,0){\includegraphics{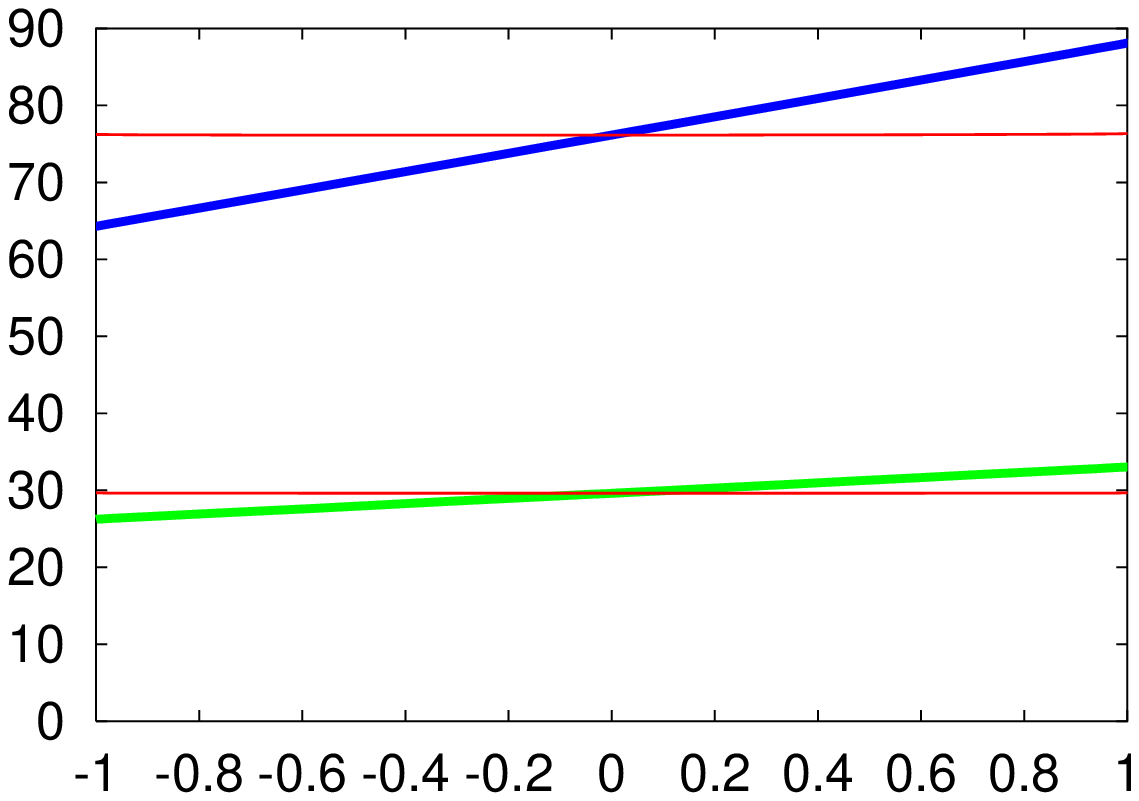}}
\put(2.3,5.1){b) Scenario~B}
\put(6.1,-.2){\small $ \cos\Theta_{e}$}
\put(-.1,5.4){\small $  \frac{d\sigma_e}{d\cos\Theta_{e}}$/fb}
\put(5.9,2.4){\small $(00)$}
\put(5.9,4.3){\small $(-+)$}
\end{picture}\par\vspace{.1cm}
\end{minipage}
\caption{Lepton decay angular distribution 
at $\sqrt{s}=2 m_{\tilde{\chi}^{\pm}_1}+10$~GeV 
in a) scenario~A and  b) scenario~B
with (thick lines) and without spin correlations (thin lines) for 
unpolarized beams (00) and for $P_{-}=-85$\%, $P_{+}=+60$\% $(-+)$, 
respectively. \la{fig_2}}
\end{figure}

\begin{figure}[t]
\hspace{-1cm}
\begin{minipage}{7cm}
\begin{picture}(7,5)
\put(-.1,0){\includegraphics{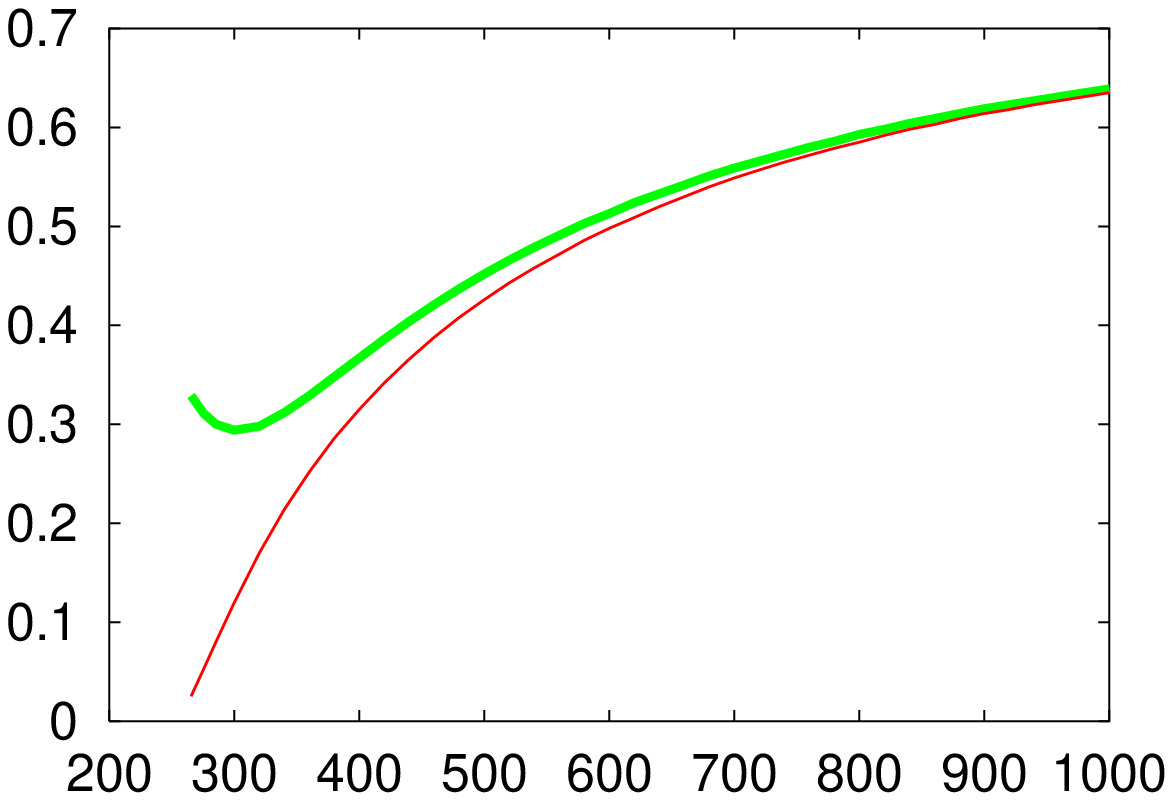}}
\put(2.3,5.1){a) Scenario~A}
\put(5.8,-.2){\small $ \sqrt{s}${\small /GeV}}
\put(-.2,5.4){\small $ A_{FB}$}
\end{picture}\par\vspace{.1cm}
\end{minipage}\hfill\hspace{.1cm}
\begin{minipage}{7cm}
\begin{picture}(7,5)
\put(-.2,0){\includegraphics{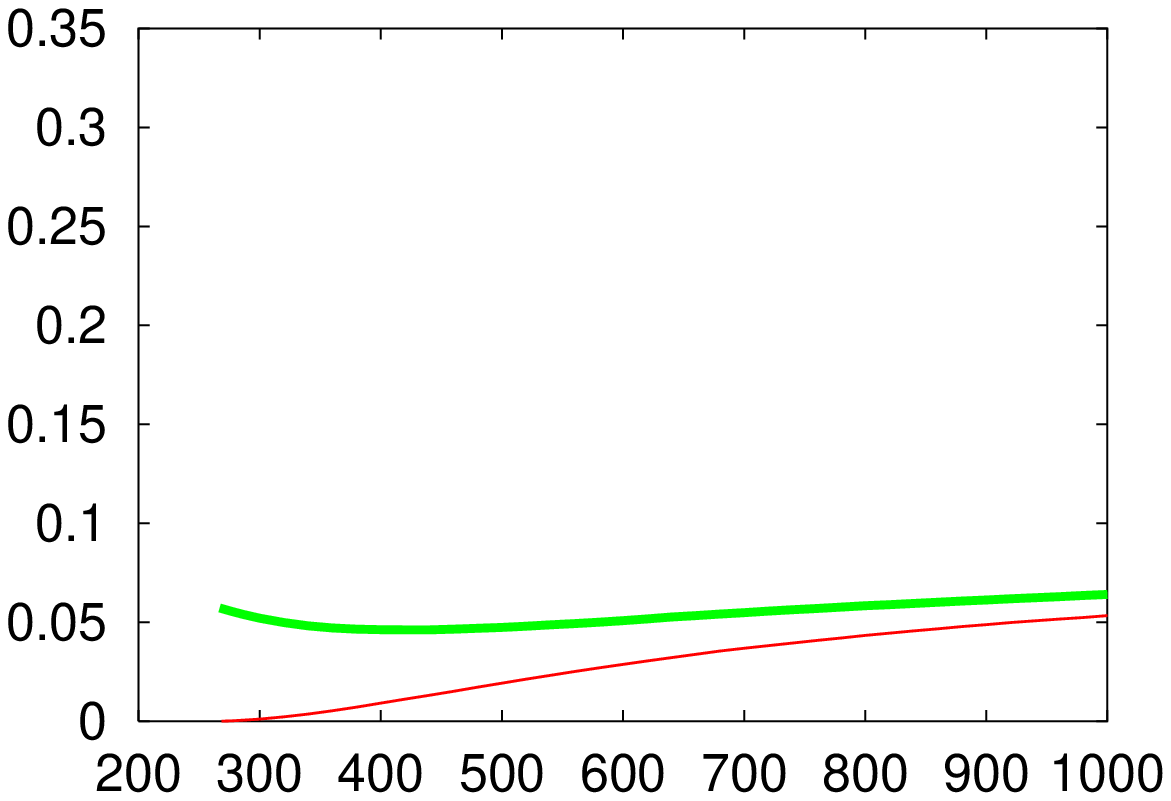}}
\put(2.3,5.1){b) Scenario~B}
\put(5.7,-.2){\small $ \sqrt{s}${\small /GeV}}
\put(-.1,5.4){\small $ A_{FB}$}
\end{picture}\par\vspace{.1cm}
\end{minipage}
\caption{Forward--backward asymmetries of the decay electron
$A_{FB}(e^+ e^-\to\tilde{\chi}^+_1\tilde{\chi}^-_1, \tilde{\chi}^-_1\to 
\tilde{\chi}^0_1 e^- \tilde{\nu}_e)$/\%
with (thick lines) and without spin correlations (thin lines)
in a) scenario~A and b) scenario~B as a function of
$\sqrt{s}$ for unpolarized beams. \la{fig_3}}
\hspace{-1cm}
\begin{minipage}{7cm}
\vspace{1.2cm}
\begin{picture}(7,5)
\put(-.1,0){\includegraphics{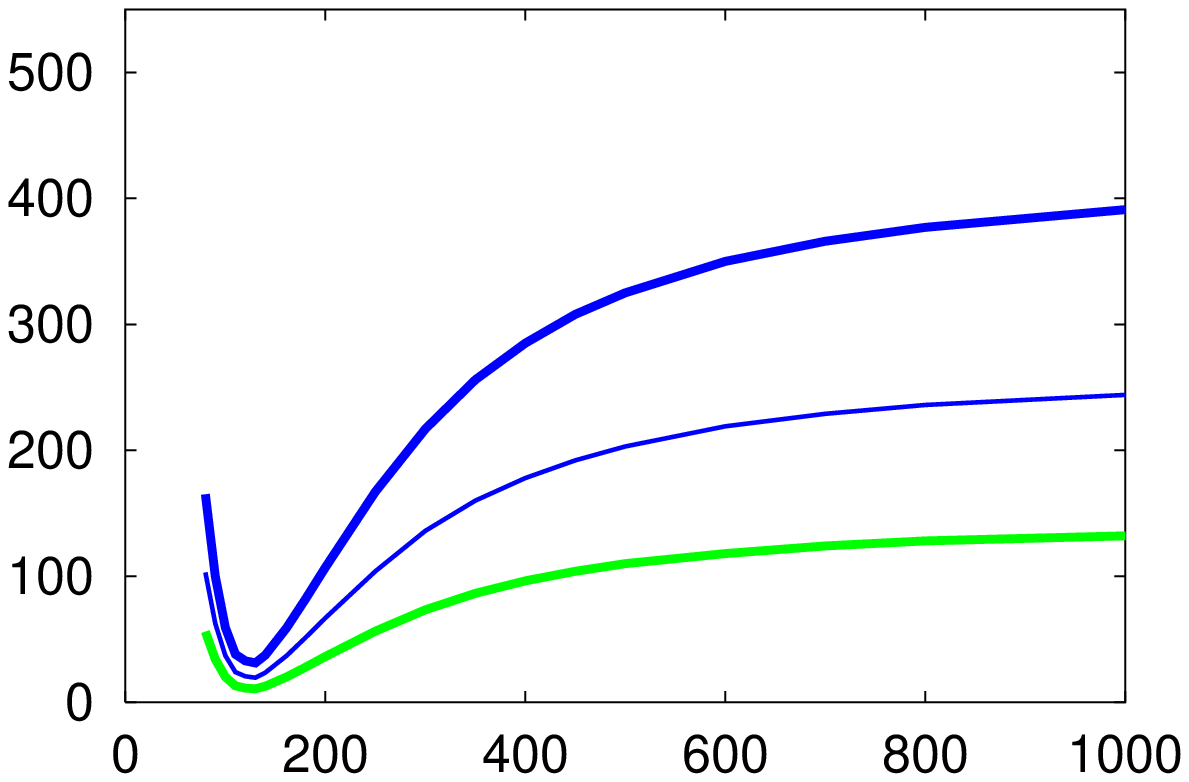}}
\put(1.8,5.1){a) $\sqrt{s}=2 m_{\tilde{\chi}^{\pm}_1}+10$~GeV}
\put(5.8,-.2){\small $ m_{\tilde{\nu}_e}${\small /GeV}}
\put(-.1,5.4){\small $ \sigma_e${\small /fb}}
\put(5.9,1.8){\small $(00)$}
\put(5.8,2.7){\small $(-0)$}
\put(5.8,3.9){\small $(-+)$}
\end{picture}\par\vspace{.2cm}
\end{minipage}\hfill\hspace{.1cm}
\begin{minipage}{7cm}
\vspace{1.2cm}
\begin{picture}(7,5)
\put(-.1,0){\includegraphics{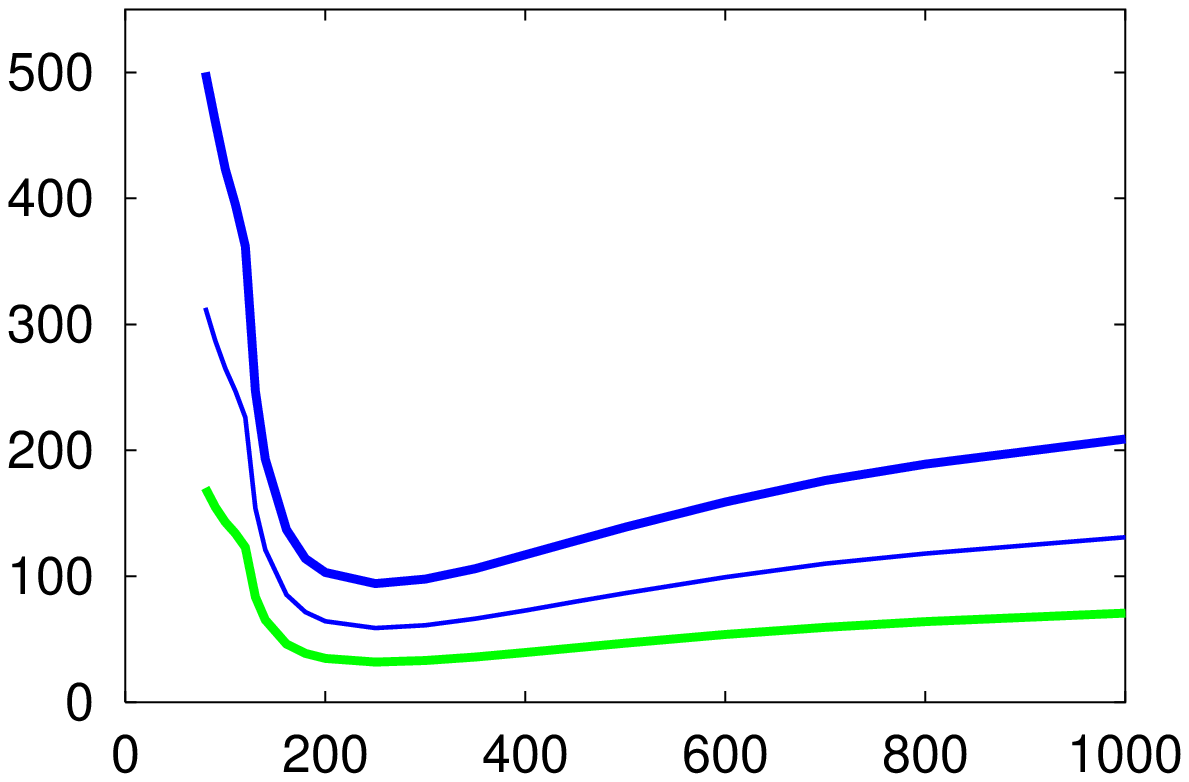}}
\put(2.4,5.1){b) $\sqrt{s}=500$~GeV}
\put(5.8,-.2){\small $ m_{\tilde{\nu}_e}${\small /GeV}}
\put(-.1,5.4){\small $ \sigma_e${\small /fb}}
\put(5.9,1.25){\footnotesize $(00)$}
\put(5.8,1.7){\small $(-0)$}
\put(5.8,2.6){\small $(-+)$}
\end{picture}\par\vspace{.2cm}
\end{minipage}
\caption{ Cross Section 
$\sigma_e=\sigma(e^+e^-\to\tilde{\chi}^+_1\tilde{\chi}^-_1)
\times BR(\tilde{\chi}^-_1\to \tilde{\chi}^0_1 e^- \bar{\nu}_e$
in a) at $\sqrt{s}=2m_{\tilde{\chi}^{\pm}_1}+10$~GeV 
and in b) at $\sqrt{s}=500$~GeV
as function of 
$m_{\tilde{\nu}_e}$ for unpolarized beams (00), 
only the electron beam polarized 
$P_-^3=-85\%$ $(-0)$ and both beams polarized $P_-^3=-85\%$, 
$P_+^3=+60\%$ $(-+)$. Other parameters as in scenario~A. \la{fig_4}}
\end{figure}

\begin{figure}[t]
\hspace{-1cm}
\begin{minipage}{7cm}
\vspace{1.2cm}
\begin{picture}(7,5)
\put(0,0){\includegraphics{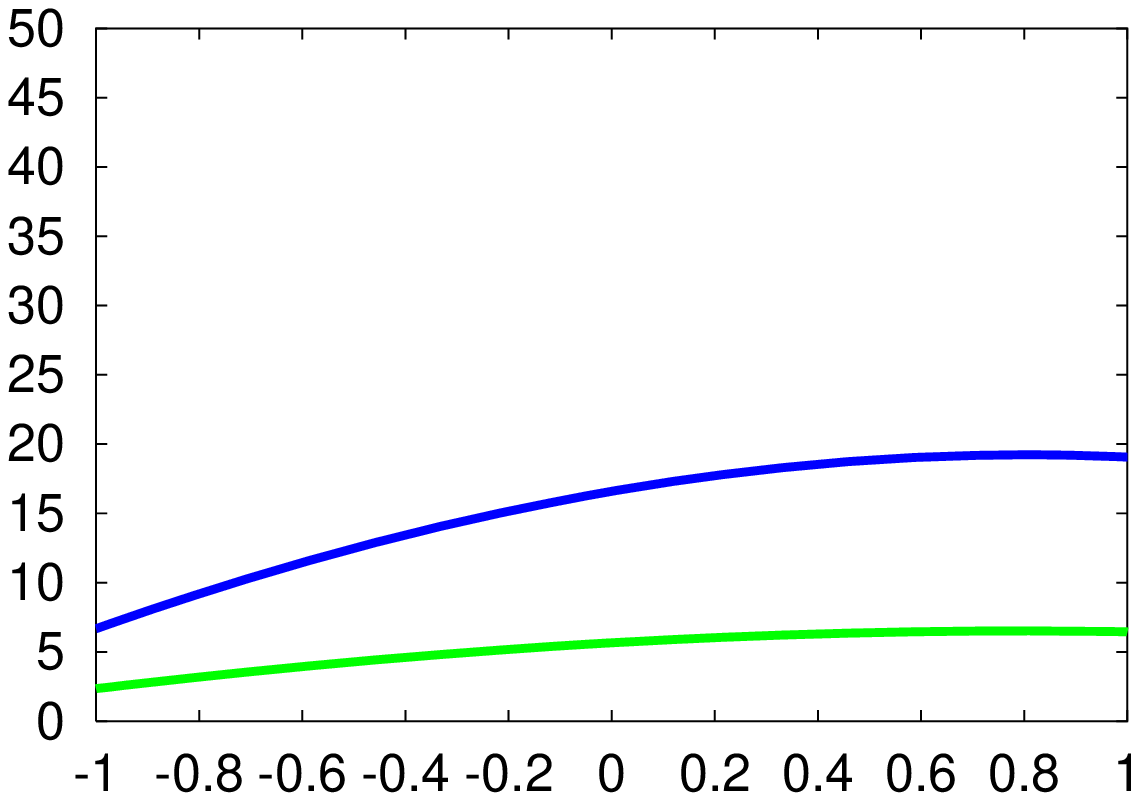}}
\put(1.8,5.1){a) $\sqrt{s}=2 m_{\tilde{\chi}^{\pm}_1}+10$~GeV}
\put(1,4.2){\small $m_{\tilde{\nu}_e}=130$~GeV}
\put(6.1,-.2){\small $ \cos\Theta_{e}$}
\put(-.2,5.4){\small $ \frac{d\sigma_e}{d\cos\Theta_{e}}$/fb}
\put(6,1.4){\small $(00)$}
\put(5.9,2.5){\small $(-+)$}
\end{picture}\par\vspace{.1cm}
\end{minipage}\hfill\hspace{.2cm}
\begin{minipage}{7cm}
\vspace{1.2cm}
\begin{picture}(7,5)
\put(0,0){\includegraphics{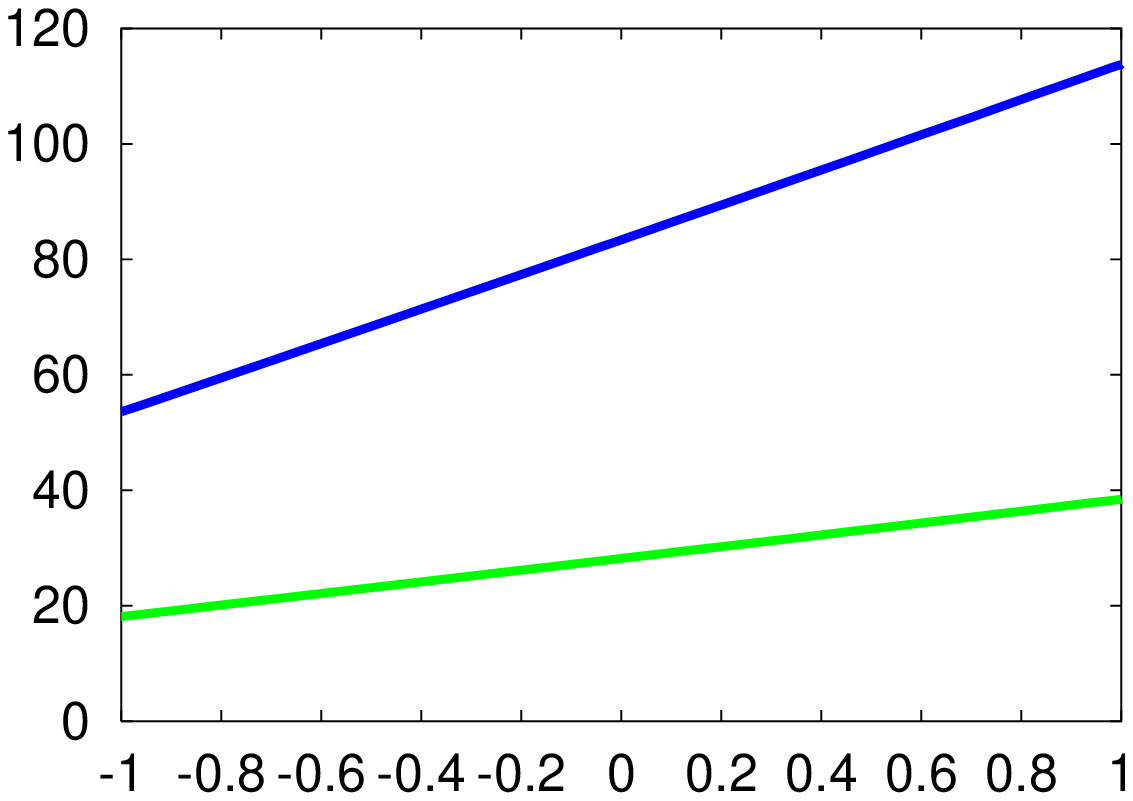}}
\put(2,5.1){b) $\sqrt{s}=2 m_{\tilde{\chi}^{\pm}_1}+10$~GeV}
\put(1.2,4.2){\small $m_{\tilde{\nu}_e}=250$~GeV}
\put(6.1,-.2){\small $ \cos\Theta_{e}$}
\put(-.1,5.4){\small $  \frac{d\sigma_e}{d\cos\Theta_{e}}$/fb}
\put(5.9,2.2){\small $(00)$}
\put(5.9,3.9){\small $(-+)$}
\end{picture}\par\vspace{.1cm}
\end{minipage}
\caption{Lepton decay angular distribution at 
$\sqrt{s}=2m_{\tilde{\chi}^{\pm}_1}+10$~GeV  
for a) $m_{\tilde{\nu}_e}=130$~GeV and
b) $m_{\tilde{\nu}_e}=250$~GeV
for unpolarized beams (00) and 
for $P_{-}=-85$\%, $P_{+}=+60$\% $(-+)$. The other parameters as in 
scenario~A but $m^2_{\tilde{e}_L}=m^2_{\tilde{\nu}_{e}}-m_W^2 \cos 2 \beta$.
\la{fig_6}}
\hspace{-1cm}
\begin{minipage}{7cm}
\vspace{1.2cm}
\begin{picture}(7,5)
\put(-.2,0){\includegraphics{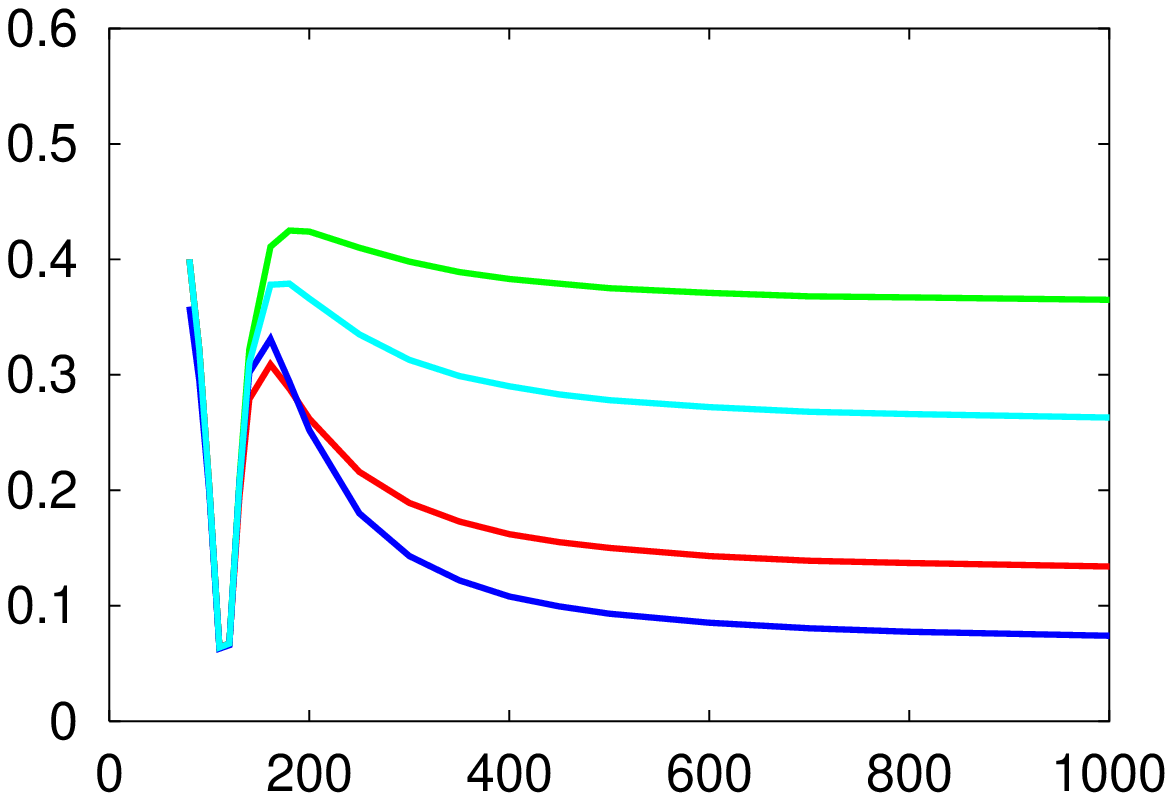}}
\put(1.8,5.1){a) $\sqrt{s}=2 m_{\tilde{\chi}^{\pm}_1}+10$~GeV}
\put(3.1,4.4){\scriptsize $P_-^3=-85\%$, $P_+^3=+60\%$}
\put(5.6,-.2){\small $ m_{\tilde{\nu}_e}${\small /GeV}}
\put(-.1,5.4){\small $ A_{FB}$}
\put(4,3.5){\scriptsize $m_{\tilde{e}_L}=130$~GeV}
\put(4,2.7){\scriptsize $m_{\tilde{e}_L}=150$~GeV}
\put(4,1.8){\scriptsize $m_{\tilde{e}_L}=200$~GeV}
\put(3,.85){\scriptsize $m_{\tilde{e}_L}^2=m_{\tilde{\nu}_e}^2-m_W^2\cos 2\beta$}
\end{picture}\par\vspace{.1cm}
\end{minipage}\hfill\hspace{.1cm}
\begin{minipage}{7cm}
\vspace{1.2cm}
\begin{picture}(7,5)
\put(-.2,0){\includegraphics{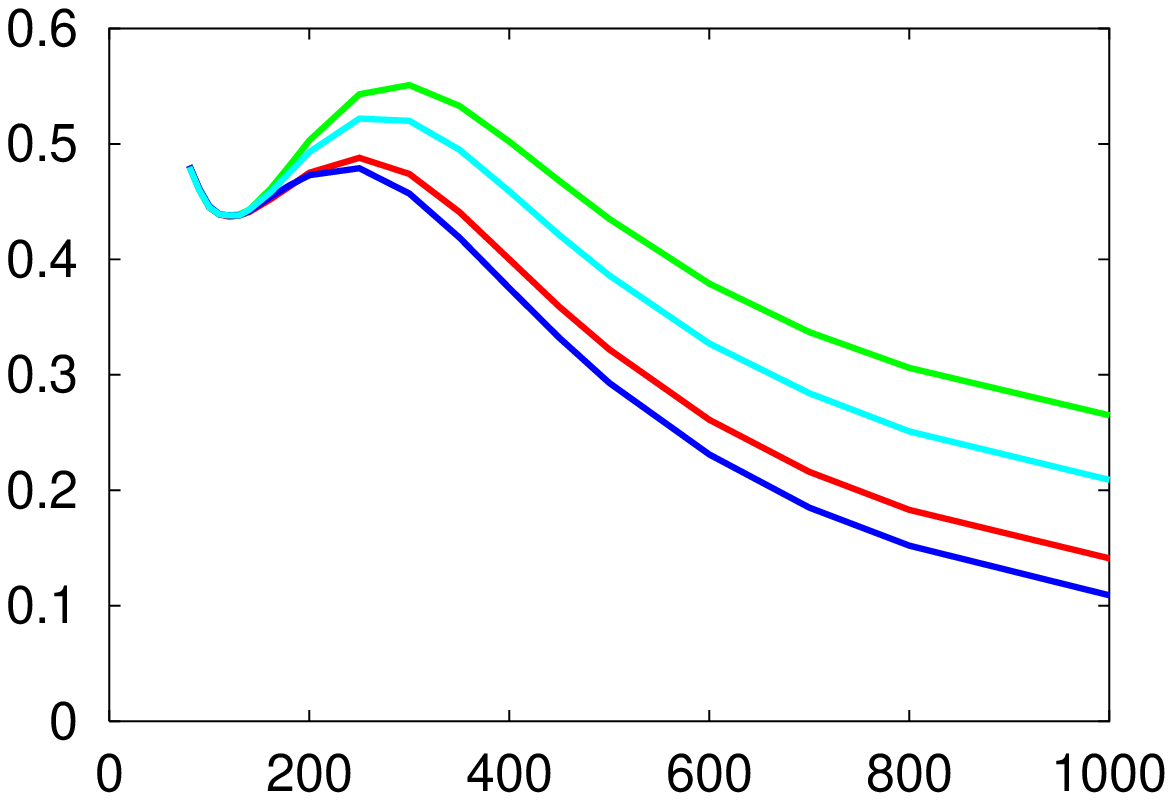}}
\put(2.4,5.1){b) $\sqrt{s}=500$~GeV}
\put(3.2,4.4){\scriptsize $P_-^3=-85\%$, $P_+^3=+60\%$}
\put(5.6,-.2){\small $ m_{\tilde{\nu}_e}${\small /GeV}}
\put(-.1,5.4){\small $ A_{FB}$}
\end{picture}\par\vspace{.1cm}
\end{minipage}
\caption{ 
Forward--backward asymmetry of the decay electron
$A_{FB}(e^+ e^-\to\tilde{\chi}^+_1\tilde{\chi}^-_1, \tilde{\chi}^-_1\to 
\tilde{\chi}^0_1 e^- \tilde{\nu}_e)$/\% in a)
at $\sqrt{s}=2 m_{\tilde{\chi}^{\pm}_1}+10$~GeV 
and in b) $\sqrt{s}=500$~GeV
as function of 
$m_{\tilde{\nu}_e}$ for 
$m_{\tilde{e}_L}=130$~GeV,
$m_{\tilde{e}_L}=150$~GeV,
$m_{\tilde{e}_L}=200$~GeV and 
$m_{\tilde{e}_L}$ fulfilling (\re{eq_su2}), 
for 
both beams polarized $P_-^3=-85\%$, $P_+^3=+60\%$ $(-+)$. 
Other parameters as in scenario~A. \la{fig_5}}
\end{figure}


\begin{thebibliography}{99}
\bibitem{bartl}  
A. Bartl, H. Fraas, W. Majerotto, Z.Phys. {\bf C 30} (1986) 441;
A. Bartl, H. Fraas, W. Majerotto, B. M\"o\ss lacher,
 Z.Phys. {\bf C 55} (1992) 257;
A. Bartl, W. Majerotto, B. M\"o\ss lacher, in '$e^+e^-$ Collisions at
500~GeV: The Physics Potential', Part B, DESY 92--123B, p. 641, ed. by 
P.M. Zerwas.\\[-1.7em]
\bibitem{Ambrosanio}  M. Chen, C. Dionisi, M. Martinez, X. Tata,
 Phys.Rep. {\bf 159} (1988) 201; S. Ambrosanio et al., in
                  {\it Physics at LEP2}, CERN 96-01, Vol.~1, p.~463,
                  eds. G. Altarelli, T. Sj\"ostrand and F. Zwirner;
                  E. Accomando et al., Phys. Rep. {\bf 299} (1998) 1.\\[-1.7em]
\bibitem{Choi} S.Y. Choi, A. Djouadi, H. Dreiner, J. Kalinowski,
               P. Zerwas, Eur. Phys. J. {\bf C 7}, 123 (1999);
               S.Y. Choi, A. Djouadi, H.S. Song, 
               P. Zerwas, Eur. Phys. J. {\bf C 8},669 (1999);
               S.Y. Choi, M. Guchait, J.Kalinowski, P.M. Zerwas,
               hep-ph/0001175;
               S.Y. Choi, A. Djouadi, M. Guchait, J. Kalinowski, H.S. Song, 
               P.M. Zerwas, hep-ph/0002033.\\[-1.7em]
\bibitem{Gudi_char}   G.~Moortgat-Pick, H.~Fraas, A.~Bartl, W.~Majerotto,
                 Eur. Phys. J. {\bf C 7} (1999) 113. \\[-1.7em]
\bibitem{Lafage} V. Lafage et al., Int. J. Mod. Phys. {\bf A14} (1999) 5075.
\\[-1.7em] 
\bibitem{Gudi_sneu} G. Moortgat-Pick, H. Fraas,
                    Acta Phys.Polon. {\bf B30} (1999) 1999.\\[-1.7em] 
\bibitem{LC99_neut} G. Moortgat--Pick, A. Bartl, H. Fraas, W. Majerotto,
                    hep-ph/0002253.\\[-1.7em]
\bibitem{Haber-Kane}  H.E. Haber, G.L. Kane, Phys. Rep. {\bf 117} (1985) 
                      75.\\[-1.7em]
\bibitem{Blair}  S. Ambrosanio, G.A. Blair, P. Zerwas, ECFA-DESY
                LC-Workshop, 1998, 
                http://www.desy.de/conferences/ecfa-desy-lc98.html.\\[-1.7em]
\bibitem{Gudi_diss} G. Moortgat--Pick, PhD Thesis, W\"urzburg 1999.\\[-1.7em]
\bibitem{Feng}  J.L. Feng, M.J. Strassler, Phys. Rev. {\bf D 55}
                (1997) 1326.\\[-1.7em]
\bibitem{fpmt} J. L. Feng, M. E. Peskin, H. Murayama, X. Tata, Phys. Rev. D 52
               (1995) 1418.\\[-1.7em]
\bibitem{Tsukamoto} T. Tsukamoto, K. Fujii, H. Murayama, M. Yamaguchi, 
                    Y. Okada, Phys. Rev. {\bf D 51} (1995) 3153.
\bibitem{Ramond} S.P. Martin, P. Ramond, Phys. Rev. {\bf D 48} (1993) 
                 5365.\\[-1.7em]
\bibitem{martyn} U. Martyn, G. Blair, hep-ph/9910416,
                 to be published in the Proceedings of 4th International 
                      Workshop on Linear Colliders (LCWS 99), Sitges, 
                      Barcelona, Spain, 28 April - 5 May 1999.\\[-1.7em]
\end{thebibliography}
\end{document}